\documentclass[aps,pra,numerical,showpacs,preprint]{revtex4-1}
\usepackage[fleqn]{amsmath}
\usepackage{amssymb}
\usepackage{amsfonts}
\usepackage[dvipdfmx]{graphicx}

\begin{document}
	
\title{Pure Coulomb tensor interaction in the Dirac equation}

\author{M. G. Garcia}
\affiliation{Departamento de F\'{\i}sica, Instituto Tecnol\'ogico de Aeron\'autica,  12228-900, S\~ao Jos\'e dos Campos, SP, Brazil}

\author{S. Pratapsi}
\affiliation{CFisUC, Physics Department, University of Coimbra,
	P-3004-516 Coimbra, Portugal}

\author{P. Alberto}
\affiliation{CFisUC, Physics Department, University of Coimbra,
	P-3004-516 Coimbra, Portugal}

\author{A. S. de Castro}
\affiliation{Departamento de F{\'\i}sica e Qu{\'\i}mica,
	Universidade Estadual Paulista, 12516-410 Guaratinguet\'a, S\~ao
	Paulo, Brazil}
\date{\today}


\begin{abstract}
\noindent
In this work we compute bound solutions for particles and anti-particles
of the Dirac equation for a pure tensor radial Coulomb potential plus a constant.
We find that the binding depends on the sign the tensor constant potential,
and allows only bound solutions for a certain sign and magnitude of the $\kappa$ quantum number,
which is related to the spin-orbit coupling in the Dirac equation.
This relation is reversed for anti-particle solutions. On the other hand, the Coulomb tensor potential,
although not biding, changes the range of $\kappa$ values for which there are bound solutions.
\end{abstract}

\maketitle

\section{Introduction}
\label{Sec:Introduction}

The Coulomb and harmonic oscillator potentials are notable potentials throughout physics,
not only because they are the perhaps the most ubiquitous potentials in physical interactions but
also because they normally are related to special symmetries of the equations of motion.
This is true in classical mechanics, non-relativistic and relativistic quantum mechanics.
In this work we explore the Coulomb potential as radial tensor potential in the Dirac equation.
The harmonic oscillator counterpart (a linear radial tensor potential whose effective potential
is an harmonic oscillator) is what became known as the Dirac oscillator \cite{Mosh_S}
a relativistic spin-1/2 system in which the spin-orbit coupling was intrinsically
relativistic and gave rise to  particular energy spectra in which $E^2$ is a linear function of the orbital momentum
quantum number $\ell$, i.e., Regge trajectories.
This shows that the tensor character of the potential is clearly related to the spin-orbit properties of the Dirac
equation, as also shown in several applications where this interaction is used \cite{Mosh_S,tensor}.
In general, spin-orbit coupling plays a major role in strong interaction as well as in condensed matter physics.

In this paper we compute the energy spectrum of an effective extended Coulomb tensor potential,
coming from a Coulomb plus constant radial tensor potential. The constant part, which gives rise to an effective
Coulomb potential, is responsible for the binding, and, as in the Dirac oscillator, gives rise to energy
spectra explicitly dependent on spin-orbit coupling, in particular on whether on has spin-aligned or spin-antialigned
states.

There are several works which consider tensor potentials with a Coulomb shape, e.g.
\cite{ZarrHR, IkhFal,ShoMou} but these papers consider only the Coulomb part and also scalar and vector potentials
in spin or pseudospin configuration. It is those last potentials in these two configurations which provide biding respectively for
fermions or anti-fermions. The tensor changes
the centrifugal potential, which is dealt in those papers either exactly or approximately through an exponential function.

A recent paper by some of the authors deals also with a similar problem in the more general context of a mapping from an one-dimensional
Schroedinger equation with a Morse potential \cite{Morse_Dirac}.

The advantage of dealing with a pure tensor potential with a Coulomb and constant term, without scalar and vector potentials which would provide bound solutions, is that one
can study in detail the solutions obtained, in particular the conditions for having bound states, the spin-orbit effect on those as well as analyse the energy
spectrum, namely the degeneracies and upper and lower limits. We thus can pin-point the effect of a tensor potential of this type, which we show is able
to bind fermions and anti-fermions. This is precisely what we will do with the present study, which otherwise will
be not possible had we considered additional potentials with other Lorentz structures. Moreover, it allows us to compare the solutions obtained with those of the Dirac oscillator, with which
they share some particular features, as is shown below.

We will follow closely the formalism of ref.~\cite{Morse_Dirac} in computing the solutions. The paper is
organized as follows: following the present introduction, in sec. \ref{Sec:dirac_tensor}
we write the first-
and second-order differential equations
for the radial upper and lower wave functions of the Dirac spinor when one has only a radial tensor potential in the time-independent
Dirac equation, and then we proceed to apply these for a radial tensor potential which is a
linear combination of Coulomb-like potential and a constant.
In Section \ref{Sec:dirac_tensor_sol}
we discuss the solutions of these equations and the energy eigenvalues for bound states,
both for fermions and anti-fermions. Finally, in Section \ref{Sec:Disc_concl} we present
and further discuss the results, including plots of the spectra, and then draw the main conclusions of this work.

\vspace{3cm}

\section{Dirac equation with a pure tensor coupling of Coulomb type}
\label{Sec:dirac_tensor}

The time-independent Dirac equation for a spin 1/2 particle with mass $M$
interacting with a tensor potential $\mbox{\boldmath $\cal U$}$ is given by
\cite{Morse_Dirac}
\begin{equation}  \label{eq_Dirac}
H\Psi=E\,\Psi\ ,
\end{equation}
where the Hamiltonian is given by ($\hbar=c=1$)
\begin{equation}  \label{H1}
H=-\mathrm{i}\,\mbox{\boldmath $\alpha$}\ldotp\nabla+\beta M+\mathrm{i}%
\,\beta\,\mbox{\boldmath $\alpha$}\cdot\mbox{\boldmath $\cal
U$},
\end{equation}
$\mbox{\boldmath $\alpha$}$ and $\beta$ being the usual Dirac matrices (variables in boldface represent a 3-dimensional vector structure).
If the vector source for the tensor is radial, one has $\mbox{\boldmath
$\cal U$}=U(r)\,\mbox{\boldmath $\hat r$}$ \cite{tensor}, where $\mbox{\boldmath $\hat r$}$ is the position unit vector.
Using the Dirac spinor for central potentials
\begin{equation}
\Psi_{\kappa m}(\mbox{\boldmath $r$})=\left(
\begin{array}{c}
\displaystyle i\frac{g_{\kappa}(r)}{r} \mathcal{Y}_{\kappa m}(%
\mbox{\boldmath $\hat{r}$}) \\
\noalign{\vskip.2cm} \displaystyle -\frac{f_\kappa(r)}{r} \mathcal{Y}%
_{-\kappa m}(\mbox{\boldmath $\hat{r}$})%
\end{array}
\right) \, ,  \label{gen_spinor}
\end{equation}
we may write the radial equation for upper and lower radial functions of the
Dirac spinor $g_{\kappa}(r)$ and $f_{\kappa}(r)$ as \cite{Morse_Dirac}:
\begin{eqnarray}
\biggl[\frac{d \ }{d r}+\frac{\kappa}{r}+U(r)\biggr]g_{\kappa}(r)&=&(M+E)f_{%
\kappa}(r)\,,  \label{Eq:D1ordRadup} \\[0.5cm]
\biggl[\frac{d \ }{d r}-\frac{\kappa}{r}-U(r)\biggr]f_{%
\kappa}(r)&=&(M-E)g_{\kappa}(r)\, .  \label{Eq:D1ordRadlow}
\end{eqnarray}
Here $\kappa$ is the quantum number
\begin{equation}  \label{def_kappa}
\kappa =\left\{
\begin{array}{ccl}
-(\ell+1) & =-\left( j+1/2\right) , & \quad j=l+1/2\qquad\hbox{(aligned
spin)} \\[0.2cm]
\ell & =+\left( j+1/2\right) , & \quad j=l-1/2\qquad\hbox{(unaligned spin)}%
\end{array}%
\right. \ ,
\end{equation}
where $j$ is the total angular momentum quantum number and $\ell$ is the
orbital angular momentum quantum number of the upper component. This quantum
number is thus related to spin-orbit and from it one obtains the orbital
angular momentum of each spinor component and the total angular momentum $j$
as well.

From the first-order equations (\ref{Eq:D1ordRadup}) and (\ref%
{Eq:D1ordRadlow}) one obtains the second-order equations
for the radial functions $g_{\kappa }(r)$ and $f_{\kappa }(r)$
\begin{equation}
\biggl[\frac{d^{2}\,}{dr^{2}}-\frac{\kappa (\kappa +1)}{r^{2}}-2\kappa \frac{%
U(r)}{r}+U^{\prime }(r)-U^{2}(r)\biggr]g_{\kappa }(r)=(M^2-E^2)g_{\kappa
}(r)\,,  \label{Eq:D2ordgUGeral}
\end{equation}%
\begin{equation}
\biggl[\frac{d^{2}\,}{dr^{2}}-\frac{\kappa (\kappa -1)}{r^{2}}-2\kappa \frac{%
U(r)}{r}-U^{\prime }(r)-U^{2}(r)\biggr]f_{\kappa }(r)=(M^2-E^2)f_{\kappa
}(r).  \label{Eq:D2ordfUGeral}
\end{equation}

The terms $-2\kappa U(r)/r\pm U^{\prime }(r)-U^{2}(r)$ can be regarded as an
effective potential $-U_{eff}(r)$ in those Schr\"odinger-like equations.

Now we chose $U(r)$ to be of the form
\begin{equation}  \label{Coulomb tensor}
U(r)=\frac{a}{r}+b\ ,
\end{equation}
where $a$ and $b$ are constants. As discussed in \cite{tensor} the potential
$\mbox{\boldmath $\cal U$}$ can arise from the gradient of the time
component of the $\omega$ meson vector field $\omega^\mu$ in a theory
describing fermions interacting with vector mesons. Therefore, one can
regard the particular form of the potential (\ref{Coulomb tensor}) as
resulting from an static radial $\omega^0$ field which is a combination of a
logarithm field and a linear confining field. Actually, as we going to see
later, not surprisingly, only the linear field is able to bind the fermions.

We can now write the first-order equations (\ref{Eq:D1ordRadup}) and
(\ref{Eq:D1ordRadlow}) with the potential (\ref{Coulomb tensor}) and obtain
\begin{eqnarray}
\biggl(\frac{d \ }{d r}+\frac{\overline{\kappa}}{r}+b\biggr)g_{\overline{\kappa}}(r)&=&(M+E)f_{%
\overline{\kappa}}(r)\,,  \label{Eq:D1ordRadupCoul} \\[0.5cm]
\biggl(\frac{d \ }{d r}-\frac{\overline{\kappa}}{r}-b\biggr)f_{%
\overline{\kappa}}(r)&=&(M-E)g_{\overline{\kappa}}(r)\, .  \label{Eq:D1ordRadlowCoul}
\end{eqnarray}
where $\overline{\kappa}=\kappa+a$. One may notice immediately that $g_{\overline{\kappa}}(r)$ and $f_{\overline{\kappa}}(r)$ are interrelated by the replacements
$\kappa\to -\kappa$, $a\to -a$ ($\overline{\kappa}\to-\overline{\kappa}$), $b\to -b$ and $E\to -E$.

In the same way, from (\ref{Eq:D2ordgUGeral}) and (\ref{Coulomb tensor}) the second-order equation for $g_{\overline{\kappa}}(r)$
becomes
\begin{equation}
\biggl[\frac{d^{2}\,}{dr^{2}}-\frac{\bar{\kappa}(\bar{\kappa}+1)}{r^{2}}-%
\frac{2b\bar{\kappa}}{r}+E^{2}-M^{2}-b^{2}\biggr]g_{\overline{\kappa}}(r)=0\,,
\label{Eq:D2ordgUCoul}
\end{equation}%
and for $f_{\overline{\kappa}}(r)$ one has
\begin{equation}
\biggl[\frac{d^{2}\,}{dr^{2}}-\frac{\bar{\kappa}(\bar{\kappa}-1)}{r^{2}}-%
\frac{2b\bar{\kappa}}{r}+E^{2}-M^{2}-b^{2}\biggr]f_{\overline{\kappa}}(r)=0\,.
\label{Eq:D2ordfUCoul}
\end{equation}%

\section{Solutions}
\label{Sec:dirac_tensor_sol}

\subsection{General case}
\label{Sec:solutions} Equations (\ref{Eq:D2ordgUCoul}) and (\ref{Eq:D2ordfUCoul}) are similar to the
radial Schr\"{o}dinger equation for a particle of mass $m$ in a singular
Coulomb potential $V=Z/r+\beta /(2mr^{2})$, namely
\begin{equation}
\biggl[\frac{d^{2}\,}{dr^{2}}-\frac{\beta +l(l+1)}{r^{2}}-\frac{2mZ}{r}%
+2m\varepsilon \biggr]u(r)=0\,,  \label{Eq:D2ordgSS}
\end{equation}%
with bound-state solutions expressed by (see, e.g. \cite{dong}, \cite{nog})%
\begin{eqnarray}
\varepsilon _{nl} &=&-\frac{mZ^{2}}{2\left(n+1/2+S\right)^2}  \notag \\
&&  \label{solCOUL} \\[-4mm]
u_{nl}(r) &=&N_{nl}r^{1/2+S}e^{-m|Z|r/\left(n+1/2+S\right)}L_{n}^{\left( 2S\right)
}(2m|Z|r/\left(n+1/2+S\right)),  \notag
\end{eqnarray}%
provided that $Z<0$ and $\beta >-1/4$. Here,$\ N_{nl}$ is a normalization
constant, $S=\sqrt{\beta +(l+1/2)^{2}}$  ($S>0$), and $%
L_{n}^{\left( \alpha\right) }$ $\left( z\right) $ with $\alpha>-1$ is the generalized
Laguerre polynomial of degree $n=0,1,2,\ldots $(see, e.g. \cite{leb}, \cite%
{abramowitz}). By comparison, identifying $Z=b\overline{\kappa}/m$, $\beta+l(l+1)=\overline{\kappa}(\overline{\kappa}+1)$ or $\overline{\kappa}(\overline{\kappa}-1)$, and $\varepsilon=(E^{2}-M^{2}-b^{2})/(2m)$,
 one can write the solutions of (\ref%
{Eq:D2ordgUCoul}) and (\ref{Eq:D2ordfUCoul}) in the form%
\begin{equation}
E_{n_{g}\overline{\kappa}} =
\pm\sqrt{M^{2}+b^{2}\left[1-\left(\frac{\overline{\kappa}}{n_{g}+1/2+|1/2+\overline{\kappa }|}\right)^2\right]}  \label{solGE}
\end{equation}
\begin{equation}
g_{n_{g}\overline{\kappa}}(r) =N_{n_{g}\overline{\kappa} }(2\gamma r)^{1/2+|1/2+\overline{\kappa }|}
e^{-\gamma r}L_{n_{g}}^{\left(2|1/2+\overline{\kappa}|\right)
}\left( 2\gamma r\right)
\label{solGG}
\end{equation}%
for $\overline{\kappa}\neq -1/2$, and
\begin{equation}
	E_{n_{f}\overline{\kappa}} =
	\pm\sqrt{M^{2}+b^{2}\left[1-\left(\frac{\overline{\kappa}}{n_{f}+1/2+|1/2-\overline{\kappa }|}\right)^2\right]}  \label{solFE}
\end{equation}
\begin{equation}
	f_{n_{f}\overline{\kappa}}(r) =N_{n_{f}\overline{\kappa} }(2\gamma r)^{1/2+|1/2-\overline{\kappa }|}
	e^{-\gamma r}L_{n_{f}}^{\left(2|1/2-\overline{\kappa}|\right)
	}\left( 2\gamma r\right)
	\label{solFF}
\end{equation}%
for $\overline{\kappa}\neq +1/2$. In these solutions $b\overline{\kappa }<0$ for both cases and $\gamma=|b\overline{\kappa}|/\left(n_{g}+1/2+|1/2+\overline{\kappa}|\right)$
for $g_{n_{g}\overline{\kappa}}$ and $\gamma =|b\overline{\kappa }|/\left(n_{f}+1/2+|1/2-\overline{\kappa}|\right)$ for $f_{n_{f}\overline{\kappa}}$ respectively.
Note that $g_{n_{g}\overline{\kappa}}(0)=0$, $f_{n_{f}\overline{\kappa}}(0)=0$, $\int_{0}^{\infty}dr\,|g_{n_{g}\overline{\kappa}}|^2<\infty$ and $\int_{0}^{\infty}dr\,|f_{n_{f}\overline{\kappa}}|^2<\infty$, as a consequence of the mapping.

The number of nodes of the radial functions $g_{n_{g}\overline{\kappa}}$ and $f_{n_{f}\overline{\kappa}}$ are determined by the degree of the
generalized Laguerre polynomials $n_g$ and $n_f$ respectively. Since $n_{f}+|1/2-\overline{\kappa }|=n_{g}+|1/2+\overline{\kappa }|$, one has
\begin{equation}
\label{n_nodes}
\left.
\begin{array}{clc}
 n_{f}=n_{g}-1\ , & \quad \overline{\kappa}<-1/2 &\quad  \ (n_g\not=0)\\[0.2cm]
n_{f}=n_{g}+1\ , & \quad\overline{\kappa}>+1/2 & \quad \ (n_f\not=0)%
\end{array}%
\right. \ .
\end{equation}
%
%
The cases for which $0<|\overline{\kappa}|<1/2$ are excluded because, using the same reasoning, it would mean that $n_g$ and $n_f$ could not be both integers.

Special attention must be paid to the cases $n_{g}=0$ for $\overline{\kappa}<-1/2$ and $n_{f}=0$ for $\overline{\kappa}>+1/2$. The solution (\ref{solGG}) with $n_{g}=0$ and $\overline{\kappa}<-1/2$ implies $E=\pm M$ from (\ref{solGE}), and can be expressed as $g_{0\overline{\kappa}} =N_{0\overline{\kappa}}r^{-\overline{\kappa}}e^{-|b|r}$. Now, since $\overline{\kappa}$ is
negative, $b$ has to be positive in order to have bound states. On the other hand, because
the right-hand side of (\ref{Eq:D1ordRadupCoul}) must be zero and at the same time (\ref{Eq:D1ordRadlowCoul}) implies that $f_{0\overline{\kappa}}(r)$ would behavior as $e^{+br}$ as $r$ tends toward infinity (and therefore
could not describe a bound state) one has to set $f_{0\overline{\kappa}}(r) =0$. In turn, from (\ref{Eq:D1ordRadlowCoul}), this selects $E= +M$, and thus, for $\overline{\kappa}<-1/2$ with $n_g=0$, one has positive energy states whose spinor has a zero lower component.
Because  $g_{\overline{\kappa}}$ and  $f_{\overline{\kappa}}$ are reciprocally connected, as already commented, one can ensure that for $\overline{\kappa}>+1/2$ the solutions with $E=- M$ describe states whose spinor has a zero upper component. Since the energies (\ref{solGE}) for fixed $b$ and $M$ depend only on the ratio $|\overline{\kappa }|/(n_{g}+1/2+|1/2+\overline{\kappa}|)$ or $|\overline{\kappa }|/(n_{f}+1/2+|1/2-\overline{\kappa }|)$, these solutions have an infinite degeneracy. It is interesting to remark at this point that in the Dirac oscillator one can have an infinite degeneracy when
the sign of $\kappa$ and of the frequency are opposite \cite{str, harm_osc_prc_2004}.

The relation between $N_{n_{g}\overline{\kappa} }$ and $N_{n_{f}\overline{\kappa} }$ can be found by substituting the solutions of the second-order equations into the first-order equations. As a matter of fact, one can explicitly compute $f_{\overline{\kappa}}$ from the solution (\ref{solGG}) using
(\ref{Eq:D1ordRadupCoul}) when $E\not=-M$, giving
\begin{equation}
g_{n_{g}\overline{\kappa}}=N_{n_{g}\overline{\kappa}}(2\gamma r)^{-\overline{\kappa }}
e^{-\gamma r}L_{n_{g}}^{\left (-1-2\overline{\kappa}\right)
}\left( 2\gamma r\right)\label{ff1}
\end{equation}
\begin{equation}
f_{n_{g}\overline{\kappa}}=-\frac{N_{n_{g}\overline{\kappa}}\,{\rm sgn}(M+E_{n_{g}\overline{\kappa}})}{\sqrt{n_{g}(n_{g}-2\overline{\kappa})}}\sqrt{\left|\frac{M-E_{n_{g}\overline{\kappa}}}{M+E_{n_{g}\overline{\kappa}}}\right |}(2\gamma r)^{1-\overline{\kappa }}
e^{-\gamma r}L_{n_{g}-1}^{\left (1-2\overline{\kappa}\right)
}\left( 2\gamma r\right)\label{f1}
\end{equation}
\begin{equation}
E_{n_{g}\overline{\kappa}} =
\pm\sqrt{M^{2}+b^{2}\,\frac{n_{g}(n_{g}-2\overline{\kappa})}{(n_{g}-\overline{\kappa})^2}}  \label{solGE1}
\end{equation}
for $\overline{\kappa}<-1/2$, and
\begin{equation}
g_{n_{g}\overline{\kappa}}=N_{n_{g}\overline{\kappa}}(2\gamma r)^{1+\overline{\kappa }}
e^{-\gamma r}L_{n_{g}}^{\left (1+2\overline{\kappa}\right)
}\left( 2\gamma r\right)\label{ff2}
\end{equation}
\begin{equation}
f_{n_{g}\overline{\kappa}}=-{N_{n_{g}\bar{\kappa}}{\sqrt{(n_{g}+1)(n_{g}+1+2\overline{\kappa})}}}\,
{\rm sgn}(M-E_{n_{g}\overline{\kappa}})\sqrt{\left|\frac{M-E_{n_{g}\overline{\kappa}}}{M+E_{n_{g}\overline{\kappa}}}\right |}(2\gamma r)^{\overline{\kappa }}
e^{-\gamma r}L_{n_{g}+1}^{\left (-1+2\overline{\kappa}\right)
}\left( 2\gamma r\right)\label{f2}
\end{equation}
\begin{equation}
E_{n_{g}\overline{\kappa}} =
\pm\sqrt{M^{2}+b^{2}\,\frac{(n_{g}+1)(n_{g}+1+2\overline{\kappa})}{(n_{g}+1+\overline{\kappa})^2}}  \label{solGE2}
\end{equation}
for $\overline{\kappa}>+1/2$. In the above relations, {\text sgn}$(x)$ stands for the sign function.

One may notice that the parameter $b$ contributes
to the mass as well, so that we may define an effective mass ${M^{\ast }}%
^{2}=M^{2}+b^{2}$. On the other hand, we see that one has $n_{g}+1/2+|1/2+\overline{\kappa }|\geq |\overline{\kappa }|$
for any $n_{g}$ and $\overline{\kappa }$, meaning
that the ratio $|\overline{\kappa }|/(n_{g}+1/2+|1/2+\overline{\kappa }|)$ is always less or equal than one and the square root is always
real for any value of $n_g$, $\overline{\kappa }$ and $b$. The minimum value of the ratio is
zero, attained when $n_g\to \infty$. This means, from Eq.~(\ref{solGE}), that
\begin{equation}  \label{E_limits}
M \leq |E_{n_{g}\overline{\kappa}}| < M^* \ .
\end{equation}
Actually, the magnitude of the energy would be precisely $M^*$ when $|\overline{\kappa }|=0$, which would happen if $a$
is a positive or negative integer, since there would be a value of $\kappa$ such that $\kappa=-a$.
However, this would not be a bound solution.

Since $M^*$ is the energy upper limit of our bound system, we may regard it as having a negative
binding energy with respect to $M^*$.
However, contrary
to what happens in non-relativistic and relativistic Coulomb vector
potentials, for positive energy solutions the usual binding energy $E_{n_{g}\overline{\kappa}}-M$ is positive.
Also we remark that the the existence of bound solutions does not depend on the value of $a$, as long as $\overline{\kappa }\not=0$.

One very special feature of this tensor interaction is that the existence of bound states depends, for a certain
value of $b$, on the sign of $\overline{\kappa }$ and thus, to a certain extent (depending on the value of $a$),
on the sign of $\kappa$, or on whether one has aligned or anti-aligned spin states, as described in the previous section.
This will be further discussed in the next section.
\subsection{$b=0$}
If $b=0$, the equation (\ref{Eq:D2ordgUCoul}) is the reduced
form of a Bessel equation of order $|1/2+\overline{\kappa}|$. This equation does
not yield bound solutions, even for integer $\overline{\kappa}$.
This result is thus independent of the value of $a$.
%
%
%
\subsection{Non-relativistic limit}
If $|b|/M\ll 1$, from equation (\ref{solGE}), one gets for the binding
energy $E_{n_{g}\overline{\kappa}}-M$
\begin{equation}  \label{En_non_rel}
E_{n_{g}\overline{\kappa}}-M \sim \frac12\,\frac{b^2}{M}\left[1-\frac{\overline{\kappa }^2}{%
(n_{g}+1/2+|\overline{\kappa }+1/2|)^2}\right] .
\end{equation}
In this non-relativistic limit, $g_{n_{g}\overline{\kappa}}$ has $n_{g}$ nodes whereas $f_{n_{f}\overline{\kappa}}\sim0$, as can be seen from (\ref{f1}) and (\ref{f2}).

Most of the discussion in the previous subsection \ref{Sec:solutions} applies here as well.
We see that the binding energy is positive and its maximum value is $\frac12\,\frac{b^2}{M}$.
The restriction that $b\overline{\kappa }<0$ still holds, so the spin-orbit dependence of the interaction
still exists. Remember that for vector (as in atomic physics, see, i.e., Ref.~\cite{BraJoa}) and for scalar type potentials
\cite{afo} the spin-orbit interaction vanishes in the
non-relativistic limit. However, at least for some tensor potentials, like the linear potential in the Dirac oscillator
\cite{Mosh_S} or in the present work, this is not the case. Here the dependence on $\kappa$, and thus on spin-orbit,
remains in the non-relativistic limit.
\subsection{Anti-fermions}
The charge conjugation operation allows to obtain anti-particle states from particle states.
In Ref.~\cite{OH1+1} it is shown that the effect of the charge conjugation is to change the sign of the
tensor potential. In our case this means that the sign of $a$ and $b$ change. Moreover,
since the charge conjugation operator changes the quantum number $\kappa$ into $-\kappa$,
one can regard an anti-fermion
state for particular values of $b$ and $a$ as the (positive energy) charge-conjugated state of the negative energy
solution of Eq.(\ref{Eq:D2ordgUCoul}) with
$b\to\--b$ and $\overline{\kappa }=\kappa+a\to-\overline{\kappa }$.

The magnitude of the energy does not depend on the sign of $b$ nor on the sign of $\overline{\kappa }$
as long as $\overline{n}=n+1/2+|1/2+\overline{\kappa }|$ (identifying $n$ with $n_g$), remains the same.
Therefore, one has $E_{n\,,\overline{\kappa }}=E^c_{n'\,,-\overline{\kappa }}$ ($n\not=0$) where $E^c$ denotes the energy
of the anti-fermion and $n'$ is such that $\overline{n}'=\overline{n}$ when the sign of $\overline{\kappa }$
is reversed. The charge conjugated states of the states described above with $E=-M$, $n_f=0$, $\overline{\kappa}>1/2$
and zero upper component,
would be the states with $E=M$, $n_g=0$, $\overline{\kappa}<-1/2$ and zero lower component.
There is, however, an important change for the existence of bound states, since it depends on the sign of
$b\overline{\kappa }$: the sign of $\overline{\kappa }$ for which one has bound states
is the opposite of the corresponding sign for particle states.
%
%
\section{Discussion and conclusions}
\label{Sec:Disc_concl}
We analyse now further the solutions uncovered in the last section.

In Fig.~1 we plot the positive energy spectrum for $a=0$ for the scaled energy
\begin{equation}
\frac{E_{n_g\overline{\kappa}}}M =
\sqrt{1+\frac{b^{2}}{M^2}\left[1-\left(\frac{\overline{\kappa}}{n_g+1/2+|1/2+\overline{\kappa }|}\right)^2\right]}  \label{solGE_esc}
\end{equation}
\begin{figure}[!ht]
\label{Fig1}
\begin{center}
\includegraphics[width=8.5cm]{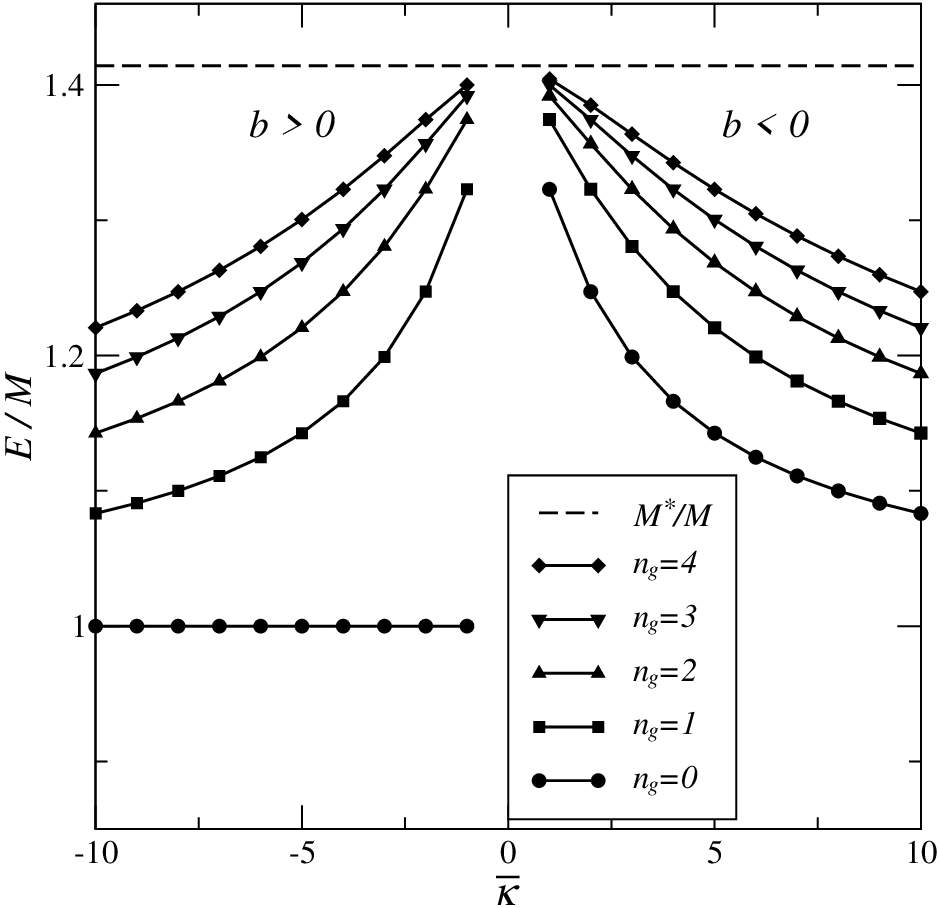}
\caption{Scaled positive energy $E/M$ for the first 5 bound state levels with $|b/M|=1$ and $a=0$.
The dashed line represents the maximum energy $E=M^*$.}
\end{center}
\end{figure}
One can see immediately that the particular case of the special solution with $\overline{\kappa }<0$ and $n_g=0$.

\vskip-.4cm
\begin{figure}[!h]
\label{Fig1}
\begin{center}
\includegraphics[width=8.5cm]{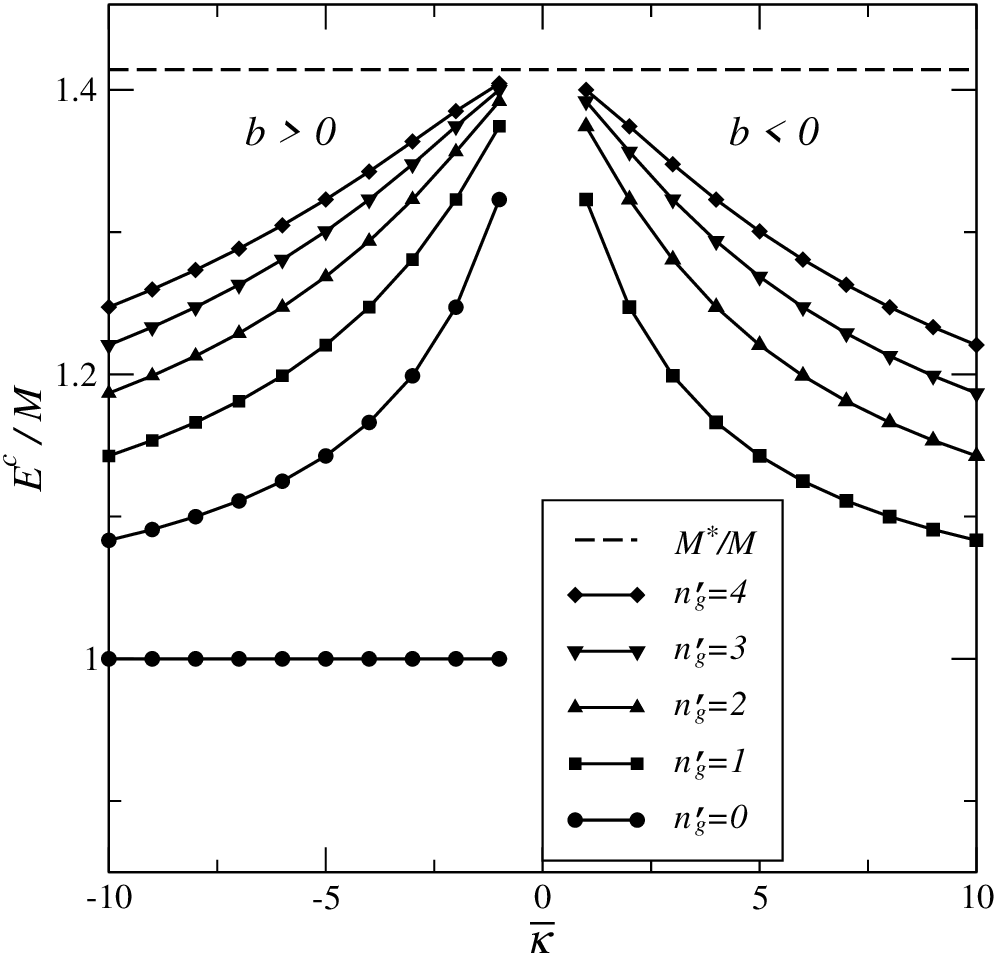}
\caption{Scaled conjugated energy $E^c/M$ for the first 5 bound state levels with $|b/M|=1$ and $a=0$.
The dashed line represents the maximum energy $E^c=M^*$.}
\end{center}
\end{figure}

One the other hand, the charge-conjugated spectrum obtained from the negative energy solution is displayed in Fig.~2.
Comparing this figure with figure 1, we see that the relation $E_{n\,,\overline{\kappa }}=E^c_{n'\,,-\overline{\kappa }}$ discussed in the previous
subsection is manifest. This spectrum can be thought as the charge conjugated spectrum from the negative energy spectrum for the parameters of Fig.~1.
As discussed in the previous section, the special solution with $E=M$ appears here as the charged conjugated solution for negative energy case when
$\overline{\kappa}>1/2$, $b<0$, with $n_f=0$.


\begin{figure}[!ht]
{
\parbox[!ht]{8.2cm}{
\begin{center}
\includegraphics[width=8cm]{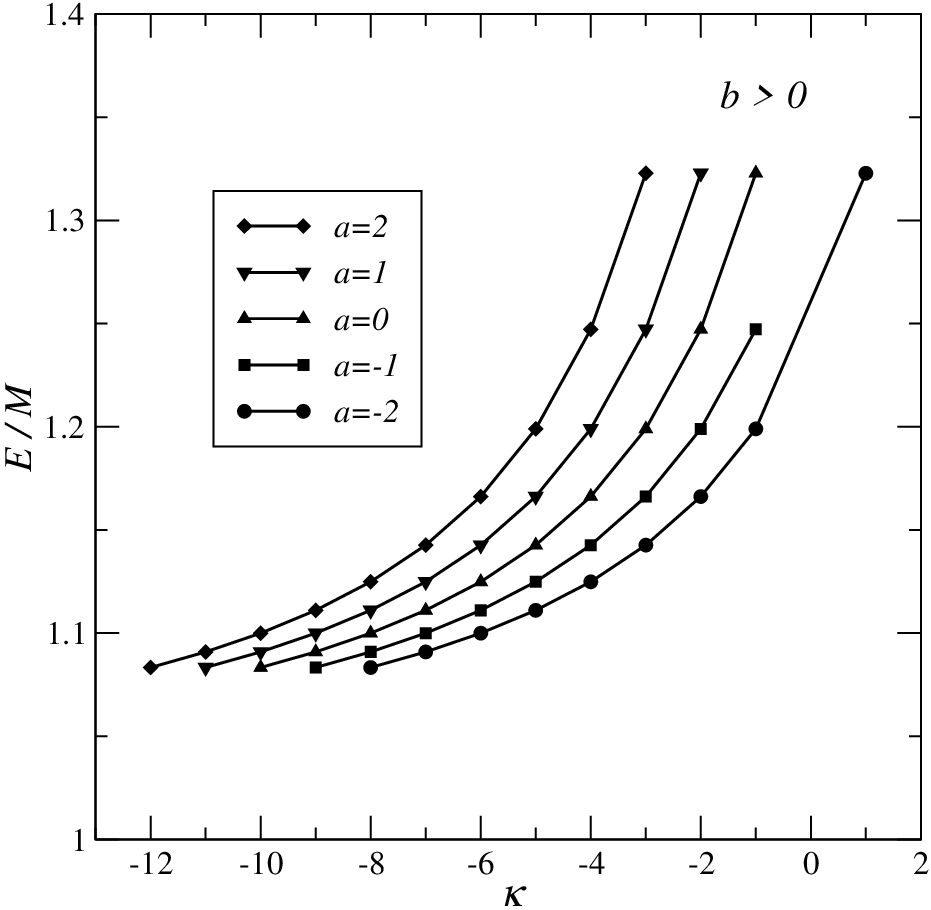}
\end{center}
}\hfill
\parbox[!ht]{8.2cm}{
\begin{center}
\includegraphics[width=8cm]{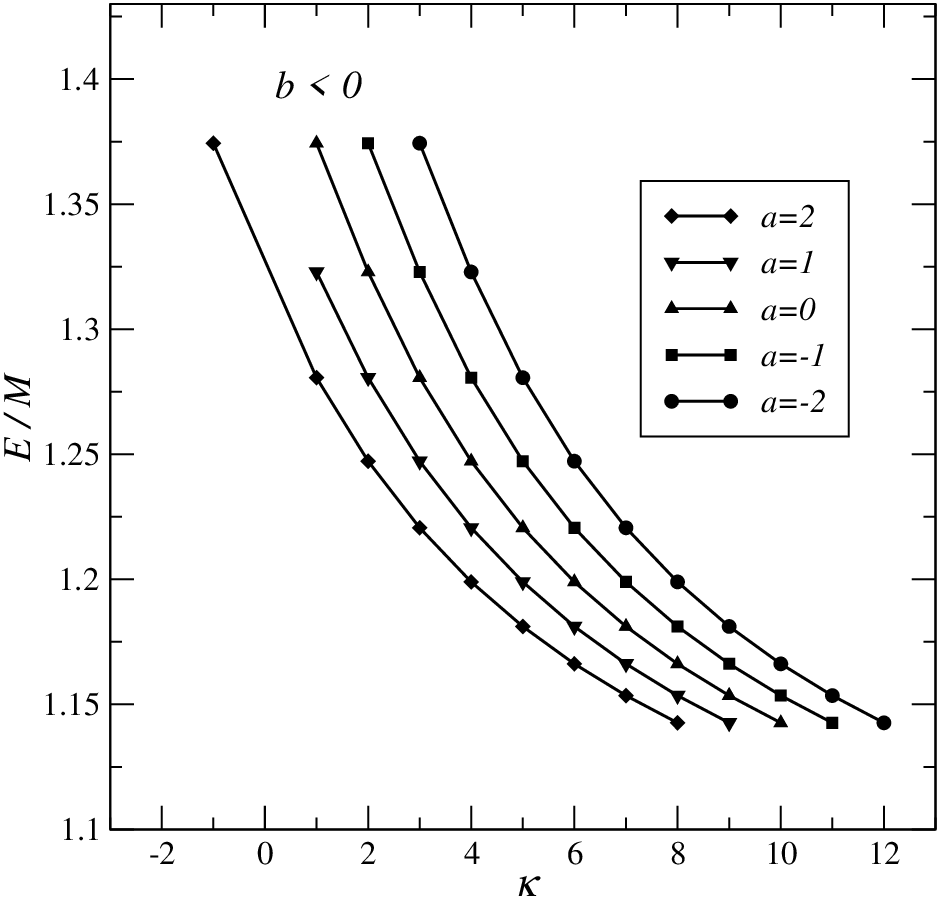}
\end{center} }  } \vspace*{-.2cm}
 \caption{Scaled positive energy $E/M$ for $n_g=1$ as function of $\kappa$ and values of $a$ ranging from $-2$ to $2$. In the left plot
 one has $-10\leq\overline{\kappa}\leq -1$ and in the right plot $1\leq\overline{\kappa}\leq 10$. The value $\kappa=0$ is excluded.}
\end{figure}

The change of the value of $a$ does not change qualitatively the spectra. One sees that from the relation $\overline{\kappa}=\kappa+a$, such that the conditions
for $\kappa$ in the two kinds of solutions, $\overline{\kappa}<-1/2$ ($b>0$) or $\overline{\kappa}>1/2$ ($b<0$), are
\begin{equation}
\label{kappa_cond}
\left.
\begin{array}{cl}
\kappa< -a-\frac12 & \quad \overline{\kappa}<-1/2\ , \quad b>0\\[0.2cm]
\kappa> -a+\frac12  & \quad\overline{\kappa}>+1/2\ , \quad b<0%
\end{array}%
\right. \ .
\end{equation}
This means that the spectra of Figs.~1 and 2 shifts horizontally if the energy is plotted against $\kappa$ as the value of $a$ changes, and the value of the energies
will change between two consecutive dots in those plots, i.e., between two consecutive integer values of $\overline{\kappa}$.
This can be seen in Fig.~3.

Although the spectra does not change qualitatively with $a$ and even the existence of bound states does not depend on it,
there is an important physical consequence of changing its value. Indeed, as one can see from (\ref{kappa_cond}),
$a$ sets the upper value of $\kappa$ when $b$ is positive and its lower value when $b$ is negative. Therefore,
while the sign of the strength $b$ of the constant tensor interaction
selects bound states with either aligned spin ($\kappa<0$) or anti-aligned spin ($\kappa>0$), adding  the
Coulomb tensor interaction whose strength is $a$ can extend or shorten the range of values of $\kappa$ for which one has bound states.
In particular, from (\ref{kappa_cond}), if the magnitude of $a$ is large enough and its sign is opposite to the sign of $b$
one can have both aligned and anti-aligned spin bound states for almost any value of $|\kappa|$.
This effect can also be seen from Figure 3.

In summary, in this work we computed the bound solutions of the Dirac equation with a pure tensor (i.e., without potentials with other Lorentz structures)
with radial dependence consisting of constant potential $b$ plus a Coulomb radial potential $a/r$. These two terms give rise, respectively,
to an effective potential with a Coulomb shape and an additional potential with a
$1/r^2$ dependence which combines with the centrifugal potential in the second-order differential equation for each radial functions in the general Dirac spinor
for central potentials. This is thus equivalent to solve a Schroedinger-like problem with a modified centrifugal barrier and a Coulomb potential, the
so-called singular Coulomb potential. The solutions are thus expressed in terms of a displaced spin-orbit quantum number $\overline\kappa=\kappa+a$ which enters both the centrifugal barrier
and the Coulomb potential, meaning that the existence of bound solutions are related to \textit{both}
the sign of the constant tensor potential and the sign of $\overline\kappa$. The former sets the latter, such that one only has bound solutions for a
particular sign of $\overline\kappa$. In turn, this sets the range of values of the quantum number $\kappa$, thus of the existence of bound states with aligned or
anti-aligned spin. When $a=0$ (no Coulomb tensor potential) $\overline\kappa=\kappa$ and one has either aligned or anti-aligned bound states. When $a\not=0$, one can have both
aligned and anti-aligned states, but one of them will not have the full range of orbital momenta, because there would be an upper or lower limit for the values of $\kappa$. A dependence of the energy spectrum on $\kappa$ can be found also in the Dirac oscillator, which also originates from a radial tensor potential \cite{harm_osc_prc_2004}.

Adding potentials with other Lorentz structure while at the same time still having analytical solutions is possible,
if, for instance, one adds a vector $V$ and scalar $S$ radial potentials in the spin symmetry limit, i.e., $V=S$
\cite{Morse_Dirac}, but, of course, more general configurations can be considered, thereby allowing for more realistic potentials.
Nevertheless, we believe that, by singling out just the tensor potential, we were able able to study
in more detail its properties and effects on the fermion and anti-fermion spectra, in particular its spin-orbit dependence.


\begin{acknowledgments}
This work was supported in part by means of funds provided by CNPq (grant 304743/2015-1) and Coordena\c c\~ao de
Aperfei\c coamento de Pessoal de
N\'{\i}vel
Superior (CAPES) - Brazil, Finance Code 001. PA would like to thank the Universidade Estadual
Paulista, Guaratinguet\'a Campus, for supporting his stays in its
Physics and Chemistry Department.
\end{acknowledgments}

\end{document}